**Title:**
**Human-agent discovery of reconfigurable in-plane ferroelectric superdomain control**

**Authors:**
Yu Liu[1*], Boris Slautin[1], Ching-Che Lin[2,3], Jaegyu Kim[3,4], Lane W. Martin[2,3,5], and Sergei V. Kalinin[1*]

[1] Department of Materials Science and Engineering, University of Tennessee, Knoxville, Tennessee 37996, USA
[2] Department of Materials Science and NanoEngineering, Rice University, Houston, Texas 77005, USA
[3] Rice Advanced Materials Institute, Rice University, Houston, Texas 77005, USA
[4] Department of Materials Science and Engineering, University of California, Berkeley, Berkeley, California 94720, USA
[5] Departments of Chemistry and Physics and Astronomy, Rice University; Houston, Texas 77005, USA

* Corresponding author: yliu206@utk.edu, sergei2@utk.edu

**Abstract**
Automated experimentation is most effective when the observables, available actions, and objective are defined before the experiment starts, as is the case for Bayesian optimization. However, in many exploratory experiments, the variables that describe the sample must be extracted from the data, new operations emerge during the experiments, and the instrument budget is too small to learn the problem by trials. Here we introduce the Scanning Probe Agentic Research Cycle (SPARC) framework, in which a coding agent and a human operator share one microscope, one notebook, and two persistent memory files. FINDINGS.md stores graded conclusions about the experiment, whereas PITFALLS.md records learned failure modes of analysis and instrument. We apply SPARC to reconfigure the in-plane superdomain direction of a (111)-oriented $PbZr_{0.2}Ti_{0.8}O_3$ film. In an operator-supervised campaign, the agent reanalyzed earlier manual measurements, defined the experimental state, and developed an oriented lattice of stationary bias pulses with alternating polarity between rows to reconfigure the superdomain direction. In a subsequent agent-controlled campaign, PITFALLS.md entries were compiled into checks that validate a design before any write. The experiments showed that spatial polarity alternation, instead of the exact matching between the lattice and lamellar periods, determines directional selection. Combining a raster scan with a masked pulse lattice printed the letters "UTK" into the superdomain orientation. The campaign also identified practical requirements for agentic experimentation: physical verification of instrument execution, the conditions under which stored findings remain valid, validation of new observables on instrument data, and robust control protocols. These results position agents as a reasoning layer for the regime before conventional optimization becomes well posed, rather than as a replacement for optimization once the experiment is already defined.

## 1. Introduction

Automated experimentation is becoming a new and distinct way of operating scientific instruments.[1] Combination of programmable instruments, high-throughput data analysis, and sequential-decision algorithms can plan a campaign as it runs, so that the experiment responds to the material adaptively instead of executing a fixed script.[2, 3] Since 2020, Bayesian optimization based on Gaussian processes, deep-kernel learning, and related methods are being rapidly adopted across automated chemical synthesis, physical depositions, and materials characterization. In scanning probe microscopy (SPM), these approaches have selected spectroscopy locations, discovered structure-property relationships, optimized resonance-enhanced measurements, and tuned microscope operation.[4-8] Related approaches have been demonstrated in high-resolution scanning transmission electron microscopy and other characterization platforms.[9, 10] Hypothesis learning goes one step further: instead of maximizing a single response, the experiment is used to decide between competing physical models, with the reduction of predictive uncertainty as the quantity being optimized.[11-14] All of these methods, however, require three things to be formalized before the experiment starts. The first is the state representation, i.e. the variables used to describe the experimental space. The second is the action parameterization, i.e. the knobs the algorithm may turn. The third is the reward, the number it is asked to maximize.

Unfortunately, many exploratory experiments do not readily allow for such structure. The variables that matter may not be known at the beginning and must be defined based on data during the experiment. In this process, a parameter initially regarded as essential may prove irrelevant, and a working experimental control primitive can be invented only after several negative results. The response can depend on instrument and sample states and local histories that were not part of the original description. In this regime the task is not simply to optimize $x$ in a discoverable or known $f(x)$, but to determine what constitutes $x$, what should be measured, and which actions are scientifically meaningful. The number of experiments is also small. A single write-and-image cycle in SPM takes tens of minutes, can irreversibly modify the sample, change the probe, hence affecting the initial state for every experiment that follows. Planning by simulating many candidate action sequences, as digital twins do,[15] needs a model of the instrument-sample system that can be evaluated cheaply. In exploratory physical science such a model is either absent when the campaign starts or has to be calibrated at length first.[16] Bayesian optimization remains valuable once a subproblem becomes well posed, but it does not by itself build the vocabulary of measurable states and admissible actions or the state-and-action grammar required to cast the problem as Bayesian optimization or as a more general dynamic decision problem.

Language-model agents provide a complementary reasoning layer for this open regime because they can work across literature, images, analysis code, instrument metadata, failed experiments, and written scientific hypotheses. LLM-based systems have translated a scientist's stated intention into executable microscope code and image analysis,[17] planned and routed multi-step atomic force microscopy (AFM) experiments across an inventory of tools,[18] and, on transmission electron microscopes, synchrotron beamlines, and robotic characterization stations, taken over sample alignment, acquisition settings, and the first pass of interpretation so that the scientist decides what to measure.[19-22] More general scientific-agent systems now automate software research, generate and rank hypotheses, or close parts of wet-laboratory and medical reasoning loops.[23-26] Most of them, however, start from a fixed workflow or from a predefined evaluator. The difficult case for physical science is the experiment that changes its own representation: the agent must decide whether a null result reflects the material or an artifact of the hardware, whether a measurement is trustworthy, and whether the next useful action is a parameter change or a qualitatively new operation.

Here, we introduce SPARC, the Scanning Probe Agentic Research Cycle, a framework in which a general-purpose coding agent and a human operator work on one microscope through a shared notebook and two persistent memory files (Fig. 2). We apply it to a model problem in which the useful action is not known at the outset, namely the control of the in-plane superdomain direction in a (111)-oriented $PbZr_{0.2}Ti_{0.8}O_3$ thin film. For conventional tip-induced ferroelectric domain writing,[27-31] the control variables are tip bias and dwell time, and the objective can be the switched area measured by piezoresponse force microscopy (PFM). For this setting, Bayesian optimization can find the optimal control conditions in a few tens of tests. Here, however, the relevant response was instead the redistribution among three in-plane polarization families (Fig. 1). The probe could also apply point force and bias pulses, pulse patterns, directional raster scans with superimposed voltage waveforms, and arbitrary trajectories with tunable bias.[32] The state representation, action space, and comparison metric were therefore unavailable at the outset and had to be developed during the experiment.

We compare two types of campaigns with different levels of human-agent interaction. In *Campaign 1* the operator executed every write while the agent constructed the state variable, the hypotheses, the controls, and the write programs, and each operator correction was recorded in FINDINGS.md or PITFALLS.md. In *Campaign 2* the agent issued the instrument commands directly, with entries from PITFALLS.md compiled into checks that validate a design before any instrument action. The materials outcome is a set of rewriting rules for controlling the in-plane superdomain direction. These rules include (i) an oriented pulse lattice with alternating polarity selects a commanded direction from the three allowed ones; (ii) a region already written to one direction can be redirected to another and retained; (iii) the polarity alternation, instead of the lattice period or a dose threshold, governs directional selection under the tested conditions; and (iv) a raster aligns virgin film along its scan direction. Combining these rules allowed to define and execute a compound operation that printed the letters "UTK" into the superdomain orientation. The interaction record and error logs also reveal the requirements and present limitations of a reliable experimental agent. The complete chronology, failure analysis, estimator studies, autonomous-loop details, prior-art analysis, and full methods are provided in the Supplementary Information.

## 2. Physics and data basis of the control problem

The control problem requires complementary physical and experimental representations. The physical representation specifies the symmetry-allowed domain states and the classes of tip perturbation that may connect them. The experimental representation defines the state variables that can be extracted from PFM images, the effects of available probe operations, and the prior knowledge supplied by earlier human-designed measurements. Keeping these representations separated is important for experiment planning since the physical model (*i.e.,* crystallography) constrains the possible domain structures and switching pathways, whereas the experimental representations (*i.e.,* measurements) establish which operations reproducibly modify those states. The latter do not, by themselves, identify the microscopic switching mechanism.

### 2.1. Physics representation and control space

We first define the physical system and associated domain hierarchy used throughout the paper (Fig. 1a). The (111)-oriented $PbZr_{0.2}Ti_{0.8}O_3$ film is tetragonal, with polarization along one of the six <001> directions. Three variants have an out-of-plane component toward the surface and three away from it. Projected onto the (111) surface, the six variants fall onto three in-plane axes. We use *domain* to denote a region containing one polarization variant and *nanodomain* for the narrow lamellae formed by an alternating pair of variants. A *superdomain* is a wider band consisting of one repeating nanodomain pair. Its in-plane director is the axis along which the lamellae run. Because a director is an unoriented axis, lateral PFM acquired at one cantilever orientation cannot distinguish its two signs. The three directors are separated by 60° in lateral-PFM images, even though the corresponding polarization-vector projections are separated by 120°. Changing the measured director means moving material between discrete superdomain families, not rotating one polarization vector continuously.[33] At the microscopic level, a change between measured directors can proceed by several routes: one or more ferroelastic steps, in which the polarization turns by 90° and the strain of the unit cell turns with it, possibly changing the out-of-plane component on the way. The change may instead arise from motion of existing walls, nucleation and growth of a competing lamellar family, or a combination of these processes (Fig. 1b). The present measurements do not distinguish these routes. At the experimental level we therefore ask a narrower question: which tip actions reproducibly redistribute the measured response among the three allowed director families?

Earlier studies of (111)-oriented $PbZr_{0.2}Ti_{0.8}O_3$ established that the stable domain configuration is a dense nano-twinned 90°-wall landscape that supports successive ferroelastic reversal, orientation-dependent coercive-field scaling, kinetically tunable multistate switching, and enhanced dielectric and pyroelectric response, with substantial contributions from domain walls.[33-37] Previous tip-induced studies in other ferroelectrics also show a strong dependence on scan path and initial state.[28, 38, 39] Which microscopic route the film takes between two directions is the subject of a companion study. Here we focus on the workflow development based on the imaging data available to the human or AI agent. We use the image as input and explore which SPM actions can reproducibly switch the film between these states. At the experimental level, we describe the domain configuration by the population vector $\boldsymbol{w} = (w_1, w_2, w_3)$, that is, the fraction of the lateral-PFM signal associated with each of the three allowed in-plane directors. Each component $w_i$ lies between 0 and 1, and the three sum to unity. The effect of a writing operation is viewed empirically as a transformation

$$\boldsymbol{w}' = F_a(\boldsymbol{w}), \qquad (1)$$

where $a$ denotes the applied tip action and $F_a(\boldsymbol{w})$ is the empirical map that action induces on the population vector. $F_a$ is a measured response function rather than a microscopic model. This description does not assume a particular microscopic switching mechanism. The same change in $\boldsymbol{w}$ may result from local polarization switching, domain-wall motion, nucleation and growth, or a combination of these processes.

PFM provides several single and compound probe operations. The tip can, for example, apply a stationary bias (Fig. 1c), follow a raster or arbitrary trajectory under DC or AC bias (Fig. 1d,e), or deliver a spatial array of discrete pulses (Fig. 1f). These operations produce different spatial distributions of electric field, stress, and strain gradient, and therefore couple differently to the three superdomain families. The relevant control variables include pulse positions and polarity, trajectory geometry and orientation, bias amplitude and waveform, and the history of the region being written. In particular, motion of a biased tip generates lateral fields and mechanical asymmetries that are known to influence ferroelastic switching.[27-30, 40] The trailing field, the field that lags behind a moving biased tip and points perpendicular to its direction of travel, is the natural first hypothesis for directional switching. The experimental workflow must relate each applied action and initial state to the observed response, test possible mechanisms, and generate new hypotheses when existing models cannot explain the observed behavior.

The complexity of the in-plane switching problem for the ferroelectric superdomains can be readily illustrated from symmetry considerations. If two domain states are related by a symmetry operation, a field that preserves that symmetry cannot preferentially select one over the other. Thus, a purely out-of-plane electric field can modify the out of plane polarization state, but by itself it does not provide a directional preference among symmetry-related in-plane variants. Selecting a particular in-plane orientation requires an additional symmetry-breaking component. That component can come from tip motion, an oriented raster or trajectory, a spatially patterned bias, or a boundary between regions written with opposite polarity. The table in Figure 1 summarizes these expectations. Uniform DC and AC biases carry no in-plane axis, whereas a moving DC-biased tip and an oriented pulse lattice do. Whether the latter operations also change the out-of-plane polarization was unknown at the start of the study. These symmetry arguments constrain possible responses but do not specify the microscopic switching pathway. We therefore combine crystallographic constraints with direct measurements of which probe operations produce reproducible transitions in this film.

**2.2. State extraction and experimental prior**

The experimental state had to be constructed from the measurements. We used lateral PFM, in which an AC bias on the tip drives a local piezoelectric shear and the torsion of the cantilever maps the in-plane response perpendicular to the cantilever axis. Lamellar superdomains therefore appear as stripe patterns whose orientation defines the director (Fig. 4a, virgin). We quantified the relative populations of three local director families from the angular power spectrum of each image, obtained by radially integrating the two-dimensional Fourier transform and binning it by angle (Fig. 4d). In the principal rewrite region, these directors were located at approximately 4°, 64°, and 124° (the 4° offset reflects the orientation of the sample relative to the image axes in the instrument coordinate system). Their 60° separation reflects the crystallographic constraint of the domain structure, whereas the overall orientation of the triad relative to the scan axes varies from one region to another. The lamellar spacing was likewise measured from the baseline image in each area and used to set the subsequent write geometry. Two images acquired with no write between them defined the local no-treatment change, or experimental floor. The director populations, lamellar spacing, and floor were therefore measured separately for each area, as discussed below. Further details are provided in Supplementary Sections S4.2 and S8.

The campaign began with an existing set of measurements acquired by the first author using fixed protocols: AC and DC raster scans, continuous trajectories, isolated pulses, and repeated imaging of written regions. Reanalysis of these data provided the empirical starting point for the agentic campaign (Fig. S1, t1–t4). The rasters had not reproducibly randomized the in-plane texture, and continuous trajectories had not produced a clear same-location change in director. These observations did not identify the switching mechanism, but they ruled out several initial strategies and motivated tests with spatially separated point pulses. A detailed reconstruction of the earlier experiments and their relation to the prior literature is provided in Supplementary Sections S1–S4.

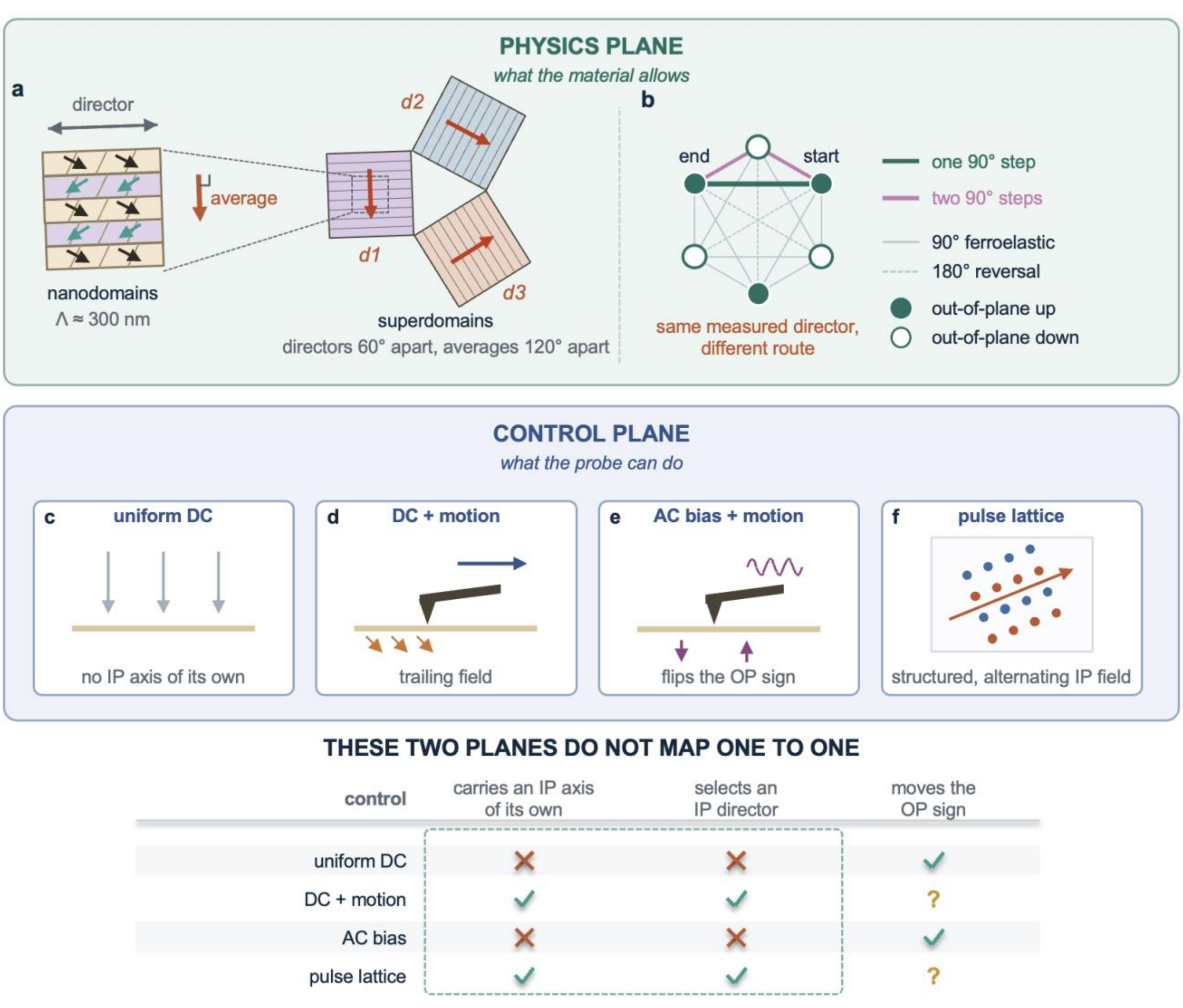

**Figure 1. The physics plane, the control plane, and the imperfect map between them. a.** A superdomain is a lamellar stack of two polarization variants. Its director is the axis along which the lamellae and their walls run. The net polarization vector shown in the schematic is normal to the walls, and crystallography allows three directors 60° apart. **b.** Variants connect by 90° ferroelastic steps or 180° reversals, and lateral PFM reads only the in-plane projection, so one change of director is consistent with more than one microscopic route. **c–f.** Four tip controls, ordered by whether the drive carries an in-plane axis of its own. A drive that preserves the 120° rotation relating the variants cannot select between them, so the table distinguishes controls that can select an in-plane director from those that cannot. Entries marked "?" were unknown at the start of the campaign.

### 3. SPARC: the scanning probe agentic research cycle

The SPARC framework pairs a general-purpose coding agent with a microscope and operator for the duration of a campaign. Five components of SPARC were developed during the study: (i) a shared notebook containing analysis and instrument-control cells; (ii) FINDINGS.md, a graded record of each conclusion together with its initial state, operation, final state, uncertainty, and any conclusion it supersedes; (iii) PITFALLS.md, which records known analysis and hardware failure modes; (iv) executable preflight checks that reject unsafe or invalid instrument actions; and (v) provenance, the record of which file, command, and instrument state produced each image, reconstructed from trajectory files, instrument headers, and frame order. The same model and tools were used throughout. The same model and tool set were used throughout. We ran two campaigns with different levels of agent authority over instrument control (Table 1). In *Campaign 1* the operator executed every write. In *Campaign 2* the agent issued instrument commands directly while the operator remained present and retained stop authority.

| **Quantity** | ***Campaign 1*** **Operator-controlled** | ***Campaign 2*** **Agent-controlled** |
|---|---|---|
| Days at instrument | 11 | 3 |
| Recorded turns | 723 | 2054 |
| Operator prompts | 68 | 117 |
| Tool calls by agent | 434 | 718 |
| Frames acquired | 70 | 155 |
| Areas used | 8 | 6 |
| Active instrument time (min) | 813 | 434 |
| Writing time (min) | not separated | 165 |
| Agent-issued closed cycles | 0 | 9 |
| Self-halts by agent | 0 | 8 |
| Graded conclusions at end | 28 | 54 |
| Rules rejected by direct test | 0 | 2 |
| PITFALLS.md gate classes executed | 0 | 5 |

**Table 1.** Census of the two campaigns. Transcript internals that depend on the harness export format are not compared. The table uses event counts and instrument records. Full chronology and counting rules are provided in Supplementary Section S5.

### 3.1. SPARC workflow and review points

Figure 2a summarizes the experimental workflow. Each iteration followed seven steps: (i) review prior knowledge from the literature, earlier measurements, and persistent memory in FINDINGS.md; (ii) generate a physical hypothesis and its predictions; (iii) plan the experiment that fixes the patterns, the doess, and the controls; (iv) review the plan against all the validation checks recorded in PITFALLS.md; (v) execute the instrument action and acquire the readout; (vi) compare the change in director populations with an untreated region in the same frame; and (vii) review the interpretation, update the memory files, and begin the next iteration. The campaigns differed in who performed the two review steps. In *Campaign 1*, the operator decided whether a proposed experiment was safe, and whether the resulting measurement was valid and physically interpretable. In *Campaign 2*, the agent made both decisions using the accumulated record and the executable checks described in Section 3.3, while the operator supplied the target and retained stop authority. The model and tool configuration were unchanged.

The record matters because it shows which decisions were the operator's and which were the agent's. Four operator questions changed the setting of the experimental campaign: how directionality should be quantified, whether the comparison should be made at the same location rather than through aggregate statistics, whether a figure in the literature had been read correctly, and whether the polarity logic of a trajectory was physically meaningful (Fig. S1, t2, t5–t8). Once a question was made precise, the agent built the estimators, generated synthetic and experimental controls, reanalyzed the earlier data, produced instrument-control files, propagated the correction through subsequent analyses, and updated the persistent record. *Campaign 1* runs are labeled R4–R7 and *Campaign 2* iterations IT1–IT10b in execution order. The full chronology is given in Fig. S1 and Table S5.

This is also what separates SPARC from agents that translate a prompt into a workflow: here the agent may change the representation that later decisions use. It can replace one state descriptor with another, redefine what constitutes a null, introduce a new action primitive, and retire a control rule when a discriminating experiment rejects it. That freedom is safe only if every revision can be traced to the data, code, and instrument state that motivated it. For this reason, the transcript, notebook, trajectory files, findings record, and raw frames are treated as parts of an integrated experimental record rather than as separate software artifacts.

### 3.2. Building persistent experimental records

The two memory files in SPARC carry information between iterations and campaigns. FINDINGS.md stores each conclusion with its start state, operation, end state, uncertainty, and a grade for how far it should be trusted. A conclusion that is later contradicted is not deleted but marked as superseded, next to the entry that replaced it. PITFALLS.md records known analysis and hardware failure modes, including anti-correlated resonance sidebands that inverted the lateral-channel sign in 10 of 46 image pairs, a void run in which the trajectory executed without bias reaching the tip, and a pulse threshold established with only one probe. The agent reread both files after each context compaction and before proposing a new design, so an operator correction entered once in *Campaign 1* was available to every subsequent iteration (Fig. S1, t6, t9, t10). Across the two campaigns, FINDINGS.md grew from 28 to 54 graded entries; 11 were later withdrawn and 3 were superseded by new measurements (Fig. 6c).

Figure 2b shows how an unexpected result was classified. When a measurement contradicted an established finding, the agent examined the probe, sample history, microscope state, and analysis before attributing the change to the material. A change in any of these factors classified the result as an artifact and added it to PITFALLS.md. If none had changed, the result was treated as a materials response, FINDINGS.md was updated, and the earlier entry was marked as superseded. The raster result of *Campaign 2* is an example of the second branch. FINDINGS.md held a graded conclusion from *Campaign 1* that a balanced raster depletes the family parallel to the scan lines. On virgin film the agent measured the opposite, an increase of the parallel family population from 0.415 to 0.601 (Fig. S5d, IT6). The probe and its contact resonance had not changed, the frame contained its own no-treatment null, and the analysis was the estimator validated in Section 4.3. The new result was therefore accepted, the earlier entry was marked superseded, and the rule was revised to depend on sample preparation. That revised rule was used later the same day to prepare the background for the patterned demonstration (Section 5.3).

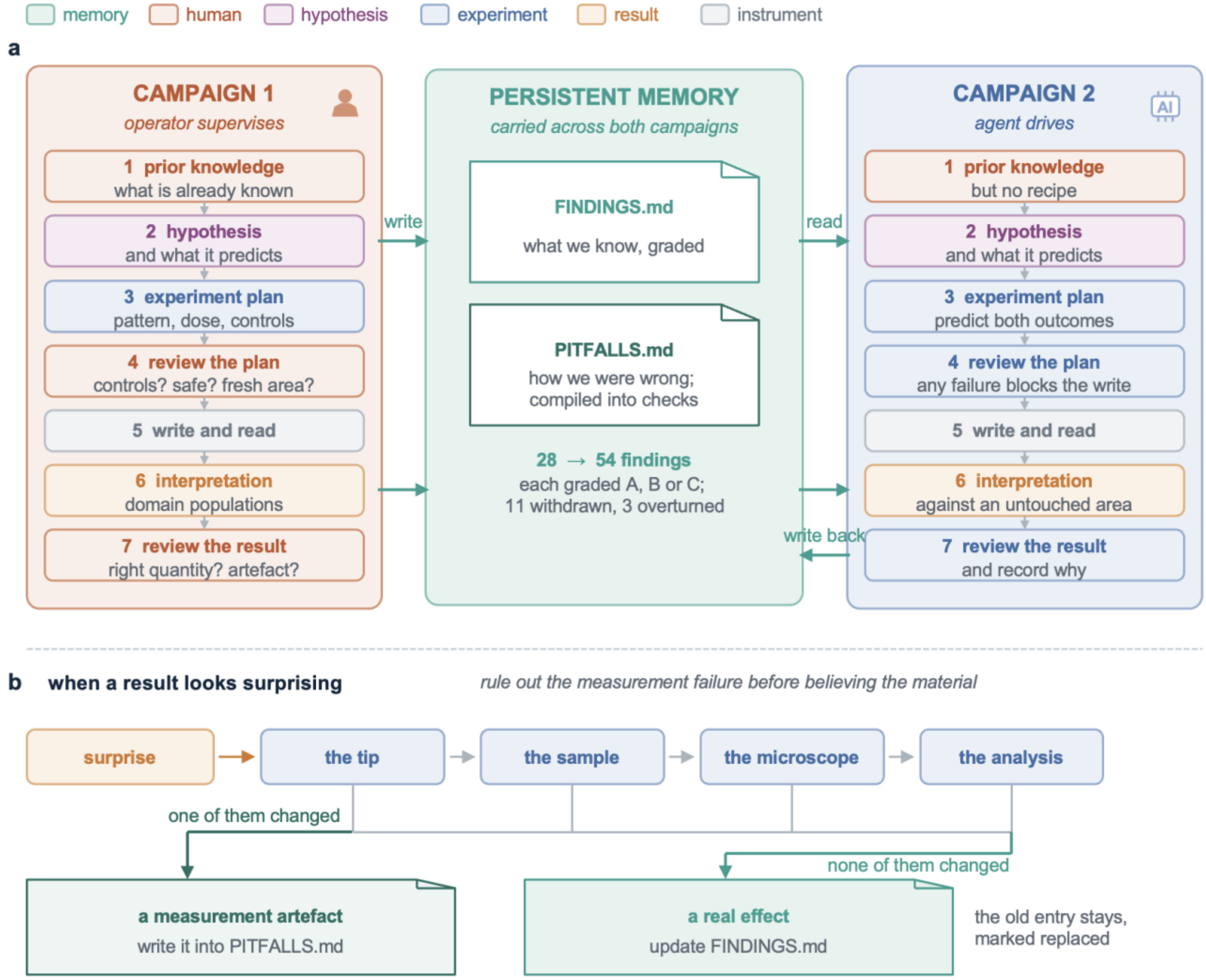

**Figure 2. The SPARC cycle, its two memory files, and the classification of an unexpected result.** **a.** Both campaigns follow the same seven steps: prior knowledge, hypothesis, experimental plan, review of the plan, instrument action and readout, interpretation, and review of the result. The campaigns differ only in who owns the two review points, steps 4 and 7. FINDINGS.md stores graded conclusions and their validity ranges, while PITFALLS.md stores known analysis and hardware failure modes, a subset of which was compiled into executable checks for *Campaign 2*. **b.** When a new measurement contradicts the record, the tip, sample, microscope, and analysis are checked before the material claim is revised. An artifact enters PITFALLS.md, whereas a reproducible physical result updates FINDINGS.md while retaining the superseded entry.

### 3.3. From recorded pitfalls to executable checks

Recorded memory alone was insufficient for agent-controlled operation because natural-language rules could still be overlooked. Direct control became practical only after selected PITFALLS.md entries were implemented as executable checks that could reject an invalid instrument action before execution. For example, the first verification safeguard addressed a void run in which the programmed trajectory completed but no bias reached the tip. The designed pulse-ladder protocol was revised to include a positive control. If neither the real pulses nor the control produced a response, the agent classified the run as a hardware failure and stopped. The agent therefore did not misinterpret the null result as insufficient voltage and respond by increasing the bias (Fig. 3c). Additional checks enforced the scanner range, bias limit, declared treatment footprint, and per-iteration instrument-action budget (Fig. S1, t14; Fig. S4a). Autonomous operation therefore depended on the accumulated record and its enforcement, not on the language model alone.

Persistent scientific memory thus became a part of the experimental infrastructure. It had to preserve a conclusion together with the evidence supporting it and the conditions under which it remained valid. Each finding was therefore linked to the initial state, applied operation, observed final state, uncertainty, and any later supersession. This structure prevented weak or obsolete interpretations from propagating silently into subsequent designs. The void run and the probe-specific threshold motivated the final form of a memory entry:

$$m = (\text{claim, evidence, uncertainty, validity conditions}) \qquad (2)$$

The need to record experimental context became clear when the agent reused a pulse threshold obtained with one probe for an experiment conducted with another. We therefore stored each threshold together with the probe and conditions used to establish it. During *Campaign 2*, the agent performed a series of hypothesis-driven, closed-loop switching experiments. Before each write–read cycle, it stated a physical hypothesis and specified the domain responses that would support or contradict it. The agent then selected a suitable region, confirmed that the planned pulse sequence remained within the predefined operating limits, applied the bias, and acquired the corresponding PFM readout. Based on the measured domain response, it either proceeded to the next experiment or terminated the campaign (Fig. S4a). Figure 3 shows representative agent outputs: a hypothesis derived from the record (Fig. 3a), selection of an observable (Fig. 3b), conversion of a pitfall into a software guard (Fig. 3c), and an audit of the control software after direct actuation was enabled (Fig. 3d). The agent-authored software still failed in several later launches; Supplementary Section S12 provides the full analysis. Here we focus on the scientific transition from an agent that developed the state variable and pulse lattice under operator supervision to one that tested those rules directly on the instrument.

### 3.4. Building the state and experimental grammar

First, the agent developed a quantitative descriptor of the domain configuration from the PFM images. It compared several methods for estimating local orientation, fitted the orientations to a three-direction model, and calculated the relative population of each direction. Successive images were then registered to compare domain changes at the same sample location (Fig. S1, t6–t8; Fig. 3b). Because different questions required different measures of the domain response, the descriptor and experimental protocol were refined together. Using this descriptor, the agent designed staged experiments around a defined initial domain configuration and a specific switching hypothesis. Each experiment included an expected response, an applied-bias condition, and an untreated control. Before testing the remaining conditions, the agent first verified that the instrument and sample produced a measurable response. The agent therefore selected measurements that tested specific hypotheses about domain switching rather than merely optimizing parameters within a fixed protocol.

**a** The agent derived the design rule the working recipe rests on, seven days before the write that used it.

> The same-sign mode leaves a standing sign template whose stripes run along the write direction, and that is what the in-plane variant selection couples to. It is the Vasudevan mechanism you corrected me on: the trajectory controls the sign/phase pattern, and the in-plane direction follows from the geometry of that pattern. Note it is also half the dose, one pass per line instead of two. Structured dose, not more dose.
>
> agent output, campaign 1, before the pulse lattice

> Two things are confounded with within_cycle and I can't separate them from n = 1: voltage went 6 to 7 V, and 90 degrees is exactly on a triad member while 0 and 60 sit midway between two.
>
> and in the same turn, unprompted

**b** Asked to compare two estimators, the agent rejected the operator's own suggestion with a structural argument, then misattributed a variance it had not measured.

> But Canny measures the wrong quantity. This is the more important finding, and it's why I wouldn't adopt it regardless of noise. Walls run parallel to stripes, so edge orientation gives the right direction, but edge length per unit area scales as area/[period]. So a finer-period family is over-counted. The control settles it.
>
> agent output, campaign 1, building the state

> The FFT is nearly unbiased but very high-variance. With only 14 to 20 periods across the region it's classic spectral-estimation variance, few independent samples per family. That explains the 0.014 to 0.047 frame-to-frame [population change] we measured on unchanged real states.
>
> four sentences earlier, the error this paper corrects

**c** After the void run, the agent converted a missing control into a check that prints a hardware verdict instead of a physical one.

> Without the positive control, a null from the ladder is ambiguous between “pulses can't reopen the gate” and “this probe or area can't switch at all”. I've changed pulse_discs so a complete null now prints a hardware/area verdict, check the bias path and coating, move at least 50 micrometres, rather than sending you up the dwell/density/repeat escalation ladder, which only applies once something has switched.
>
> agent output, campaign 1, after the void run

**d** Once it held the actuator, it replayed its own code against real frames and reported that the check encoding every safety rule had never run.

> Five defects found by simulating the real code path offline, the thing PITFALLS section 10.1 already told me to do and I hadn't been. Four had been live since IT1. IT2's written area was missing from the state file. It crashed in the verdict block before save_state, so used_areas never recorded (-14, 0), meaning S9 and check_area_fresh were both blind to it and IT4 could have written straight on top. preflight(), the S1 to S29 envelope as code, has never been called by any driver. grep -c gives 0, 0, 0 for IT1 to IT3. It crashed on prop['offset'] being None, which every driver's design guarantees, so it was unusable for exactly the case it was written for. The control tiles sat on a written panel. Audited against IT3's own frames: 3 of 14 tiles overlapped P8. Excesses move by at most 0.022 and no verdict changes, C41 stands, but that was luck.
>
> agent output, campaign 2, autonomous phase

verbatim agent output — a claim made in the same turn that later proved wrong

**Figure 3. Representative agent outputs from the SPARC cycle. a,** A design rule derived from the experimental record seven days before the instrument action in which it was used. The same exchange also contains a confound that could not be resolved with one sample and was addressed later by the exposure series. **b,** The agent rejects the operator's suggested estimator for a physically motivated reason but incorrectly attributes an unmeasured source of variance; the transfer test in Fig. S3 resolves the issue. **c,** After the void run, the missing positive control becomes an executable check that classifies the result as a hardware failure instead of increasing the exposure. **d,** After receiving direct instrument access, the agent tests its own code against real images and finds that the function intended to enforce all safety checks was not called. Verbatim agent output is shown in blue and claims later found to be incorrect in pink. Supplementary Section S6 provides the complete transcript, selection rule, and additional examples.

## 4. *Campaign 1*: Operator-controlled discovery of a reconfigurable primitive

*Campaign 1* was designed to identify a probe operation that could reorient the in-plane superdomains after earlier raster and trajectory protocols had failed. The agent did not directly control the microscope. It analyzed the evolving state, proposed and encoded instrument actions, and interpreted the results, while the operator executed the commands and retained physical control. The immediate goal was first to select one allowed direction and then to redirect a written region to another direction. Because the useful operation and its parameterization were not known in advance, the problem could not initially be formulated as a conventional Bayesian-optimization task.

### 4.1. From continuous trajectories to an oriented pulse lattice

*Campaign 1 did not follow a fixed optimization trajectory* because the useful action itself was redefined during the campaign. Reading the code that turns a requested trajectory and bias into the instrument's write file showed two things: nominally different polarities could produce identical files, and in a dense trajectory the bias reversed sign every few tens of nanometres, far inside one lamellar period. Each superdomain therefore experienced the bias flip many times over the relevant lengthscale instead of spatially uniform field. The operator proposed replacing continuous trajectories with spatially separated point pulses. Initial pulse arrays did not erase the texture, so the concept was reformulated from an eraser into a directional selector (Fig. S1, t4; Fig. 3a). The agent then constructed an oriented square pulse lattice from the locally measured director and length scale, rotated it to a commanded direction, alternated polarity between rows, and generated matched control conditions (Fig. S2b). In the principal region the measured 280 nm period gave a 140 nm row spacing. Each site received 10 V for 1 s. We quote dose as voltage-time exposure, bias multiplied by dwell time summed over sites and divided by area, because current was not measured and no charge can be assigned.

The first decisive test asked whether the pulse lattice simply printed its own geometry into the image or selected among the three directions the crystal allows. A command placed on one of the three fitted directors increased power at the commanded direction by approximately +0.385 and made that family dominant. A command approximately 26° from the nearest allowed director produced a smaller +0.152 change (R5, Fig. S2a), while the dominant response remained on a member of the triad. The experiment does not show zero response away from the allowed directions. It shows that the strongest response lands on one of the three allowed directions, not on the direction of the command. A later site-count confound does not affect this comparison because both panels contained 144 sites (Supplementary Section S7.2).

### 4.2. Reconfiguration of an already written region

The better-defined experiment (R6) asked whether an established in-plane population could be redirected. Three 2 $\mu m$ panels, P1, P2, and P3, were defined within one 8 $\mu m$ frame (Fig. 4a-c, rows), and the frame was imaged in three states: virgin, after a first pulse-lattice write, and after a second (Fig. 4a-c, columns). Write A was a lattice commanded to the 4° member of the fitted triad, and write B to the 124° member. P1 received A and then B. P2 was untreated during A and then received B. P3 received A and no second write, so that P2 is the no-treatment control for A, and P3 is the retention control for B. The fitted triad was approximately 4°, 64°, and 124°. We report the result as changes in power at the commanded direction, measured against the no-treatment floor, because those are the quantities that did not depend on which estimator was used (Section 4.3). During the first write, power near 4° in P1 increased by +0.392 and the independently written P3 increased by +0.577 while the untreated P2 control moved by −0.006, within the run-specific floor of 0.018 (Fig. 4e, g). During the second write, power near 4° in P1 decreased by -0.287 while power near 124° increased by +0.149 (Fig. 4d, e, g). The A-only retention control P3 relaxed by only -0.076 over approximately the same interval (Fig. 4c, g), and write B on the untreated P2 raised its 124° power by +0.111 (Fig. 4b, g). The instrument log confirmed that the two lattice writes were the only biased operations applied to the region between the images.

The population-vector analysis gives the same qualitative result. With the Fourier estimator, P1 changed from $\boldsymbol{w}$ = (0.23, 0.56, 0.21) to (0.76, 0.08, 0.16) after A and to (0.42, 0.14, 0.43) after B (Fig. 4d). The final state remained a mixture of two families rather than a complete reorientation. Dominant-angle labels are also unstable when two families have similar weight, and the apparent peak shift in untreated P2 during action A (Fig. 4f) illustrates this limitation. We therefore base the claim on the controlled change in population rather than on a single dominant-angle label. FINDINGS.md records the result accordingly as partial reconfiguration, together with the no-treatment and retention controls.

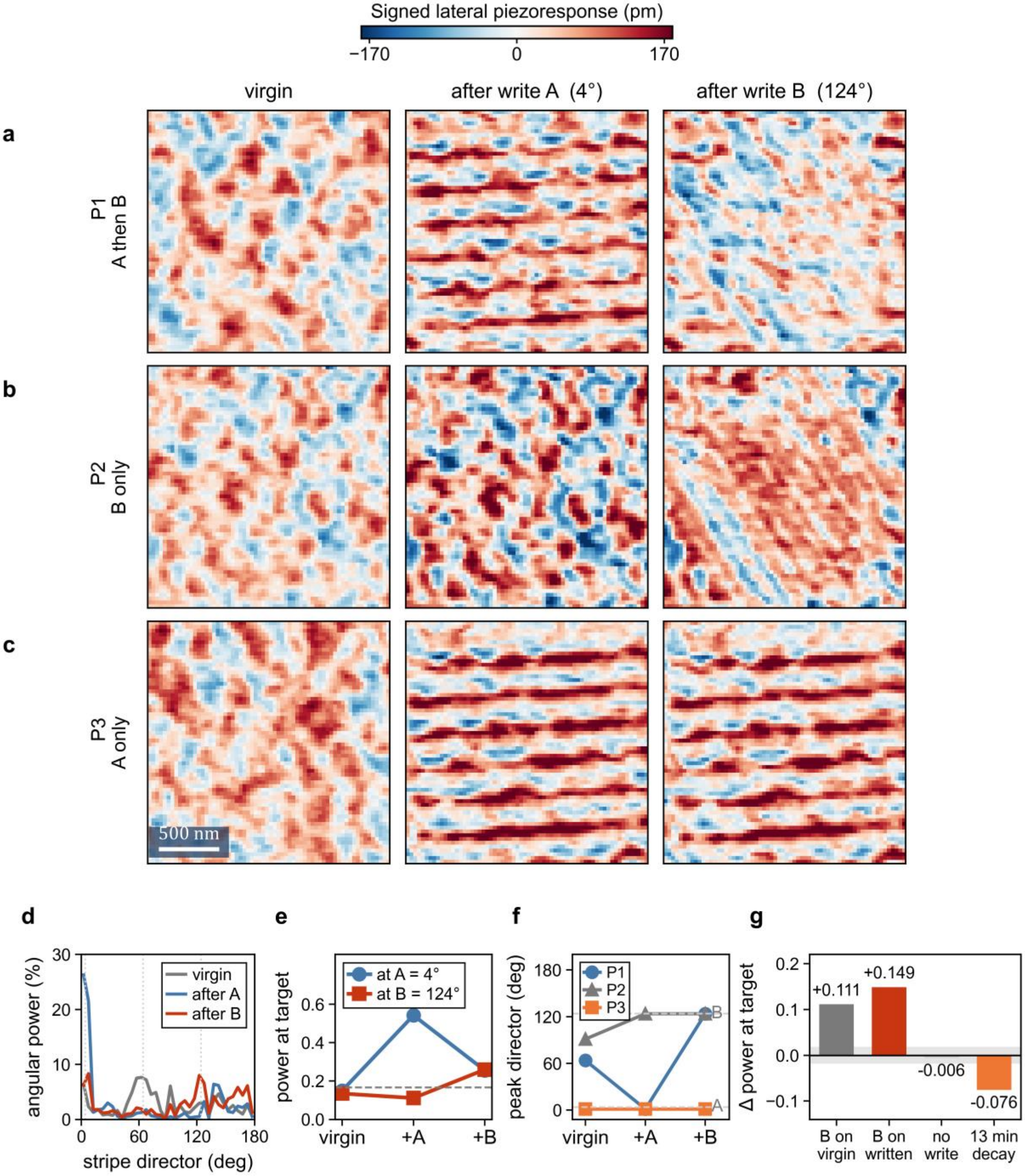

**Figure 4. Operator-supervised discovery of a rewritable in-plane superdomain direction. a–c.** Signed lateral-PFM maps of three 2 $\mu m$ panels in one 8 $\mu m$ frame, imaged in the virgin state, after write A at 4°, and after write B at 124°, on one color scale. P1 receives A and then B. P2 is untreated during A and then receives B. P3 receives A only and is the retention control. **d.** Angular power spectrum of P1 in the three states. **e.** Power at the fitted directors for P1. **f.** Dominant director of each panel. **g.** Directional contrasts relative to the measured no-treatment floor. Write A establishes the commanded family, write B redirects the P1 population toward a second family while removing most of the first, and P3 retains most of the A-written response. The final P1 state is a two-family mixture, not a complete overwrite.

### 4.3. Validation of the experimental state variable

Because the state was constructed during the campaign, its validation is part of the result. The agent compared three ways of reading stripe direction from an image, the Fourier annulus used above, an edge detector, and a structure tensor that estimates local orientation from image gradients, on synthetic stripe patterns with known ground truth. The structure tensor did best on the synthetic data (Fig. S3a), and the agent correctly identified why the edge detector fails: finer stripes have more wall per unit area, so counting edges overweights them. The ranking reversed on repeated experimental frames of an unchanged state: the Fourier estimator was about 2.5 times more repeatable (Fig. S3b). Reanalysis of the principal reconfiguration with both estimators preserved the sign of all six directional contrasts and gave strongly correlated magnitudes (Fig. S3c), although absolute populations and dominant-angle labels varied (Fig. S3d). We therefore report treatment-induced contrasts relative to an empirical null and treat absolute state labels as estimator dependent. The synthetic ranking that failed on real frames went into PITFALLS.md. The rule that survived, report changes against a measured null and name the estimator whenever an absolute value is quoted, is the one every number in this paper follows (Fig. 3b). Supplementary Section S8 gives the full transfer test.

## 5. *Campaign 2*: agent-controlled testing and composition of control rules

In *Campaign 2*, the agent operated the microscope through the Automated Experiment in Scanning Probe Microscopy interface[32] using the results and experimental rules established in *Campaign 1*. The *Campaign 2* tested two working hypotheses for preferential selection of a domain orientation. The first was that selection required the repeat spacing of the applied-polarity pattern to match the lamellar domain width. The second was that selection occurred only above a threshold electrical exposure, defined by the applied voltage and pulse duration per unit area. Before each experiment, the agent specified the PFM response expected if each hypothesis was correct or incorrect and then tested these predictions experimentally (Fig. S5).

### 5.1. Autonomous hypothesis tests

At the same electrical exposure, alternating-polarity pulse lattices increased the population of the targeted domain orientation for repeat distances of one, two, four, and eight lamellar periods (Fig. S5a, IT1–IT4). No spacing consistently produced a larger change across repeated experiments, indicating that exact matching between the lattice period and lamellar width was not required. Polarity alternation was important because uniform-polarity arrays produced no directional change above the baseline variation, whereas the alternating-polarity control clearly increased the targeted orientation (Fig. S5b). Directional selection was also observed at 70% of the exposure threshold estimated from *Campaign 1*. Moreover, the final populations of the three domain orientations remained similar over an approximately 2.2-fold exposure range (Fig. S5c, IT5). These results did not support either exact spatial matching or the proposed exposure threshold as necessary conditions (Fig. S5f). Both assumptions were therefore marked as superseded in FINDINGS.md, while polarity alternation was retained. Under the conditions tested, directional selection required alternating pulse polarity but not an exact period match or a sharply defined exposure threshold.

The response to raster writing depended on the prior writing history of the region. *Campaign 1* measurements, obtained mainly from previously written areas, suggested that a balanced raster reduced the population of the domain family parallel to the scan lines. Here a balanced raster refers to a voltage–time-balanced bipolar raster comprising one +9 V pass followed by one −9 V pass along the same scan path. In a previously unwritten region, however, the parallel-domain fraction increased from 0.415 to 0.601 after raster writing (Fig. S5d, IT6). After this result passed the checks described in Section 3.2, the agent replaced the earlier rule with a preparation-dependent interpretation. This result does not directly conflict with the lattice experiments because raster and lattice writing differ in tip motion, line continuity, and bias duty cycle. These differences prevent the contrasting responses to uniform-polarity raster and lattice patterns from being attributed to polarity alone. Instead, they indicate that the writing trajectory or temporal delivery of the bias may also affect domain selection, as examined in Supplementary Section S10.

**5.2. Set, reverse, and hold the orientation state**

*Campaign 2* next tested whether pulse-lattice writing could reversibly select a superdomain orientation and whether the written state persisted during subsequent imaging. In one region, writing increased the population of the 64° family to 0.696 from an initial state dominated by the 4° family (Fig. S5e, blue; IT9). In a second region, successive writes changed the dominant orientation and then returned it toward 64°, while an untreated reference remained unchanged (Fig. S5e, orange). One of the nine trials was excluded because the reference control failed. Of the eight valid trials, seven ended with the targeted orientation dominant, whereas one ended with another allowed orientation. The pulse lattice therefore achieved the intended orientation in seven of eight valid trials but did not provide deterministic selection. The written state persisted for 34 min during repeated imaging, but longer measurements are required to establish long-term retention.

**5.3. Composing the rules: printing UTK into the superdomain orientation**

In the final *Campaign 2* experiment (IT10b), the agent combined raster and pulse-lattice writing to produce a spatial pattern in the superdomain orientation. Both writing operations had previously been evaluated against untreated regions within the same PFM image and recorded in FINDINGS.md. Earlier measurements also showed that the written orientation persisted over the PFM readout time. A bipolar raster was first applied over an approximately 12 $\mu m$ field to produce a background dominated by the 62° family (Fig. 5a). An alternating-polarity pulse lattice oriented toward the 2° family was then applied only within a mask defining the letters 'U', 'T', and 'K' (Fig. 5b,c). The masked lattice contained equal numbers of positive and negative pulse sites. Difference maps of the 2°-family population before and after writing reproduced the letter mask (Fig. 5d–f). Relative to the spaces between the letters, the target population increased by 0.30, 0.42, and 0.52 within the U, T, and K strokes, respectively (Fig. 5g). Each letter was approximately 3 $\mu m$ across, and the effective resolution was limited by the analysis window, which had to span several lamellar periods. The experiment therefore demonstrated mask-defined patterning of the measured in-plane superdomain response, instead of arbitrary nanoscale control. Because the vertical PFM response also changed, probably during raster preparation, the complete polarization vector within the patterned regions could not be assigned. Supplementary Section S11 describes the mask construction and spatial analysis.

A moving biased tip produces an asymmetric electric-field and stress distribution behind the contact (Fig. 1d). We therefore hypothesize that a raster aligned with one of the three allowed superdomain axes favors growth of bands along that axis, consistent with reports on related (111)-oriented films.[41] The out-of-plane polarization state may also alter which in-plane variants are accessible. The present measurements do not distinguish between electrical and mechanical contributions from the moving contact or determine the role of the out-of-plane state. Resolving these possibilities will require matched rasters at several scan angles and both bias polarities, together with simultaneous out-of-plane measurements. The trailing-field explanation should therefore be regarded as a testable hypothesis rather than an established mechanism.

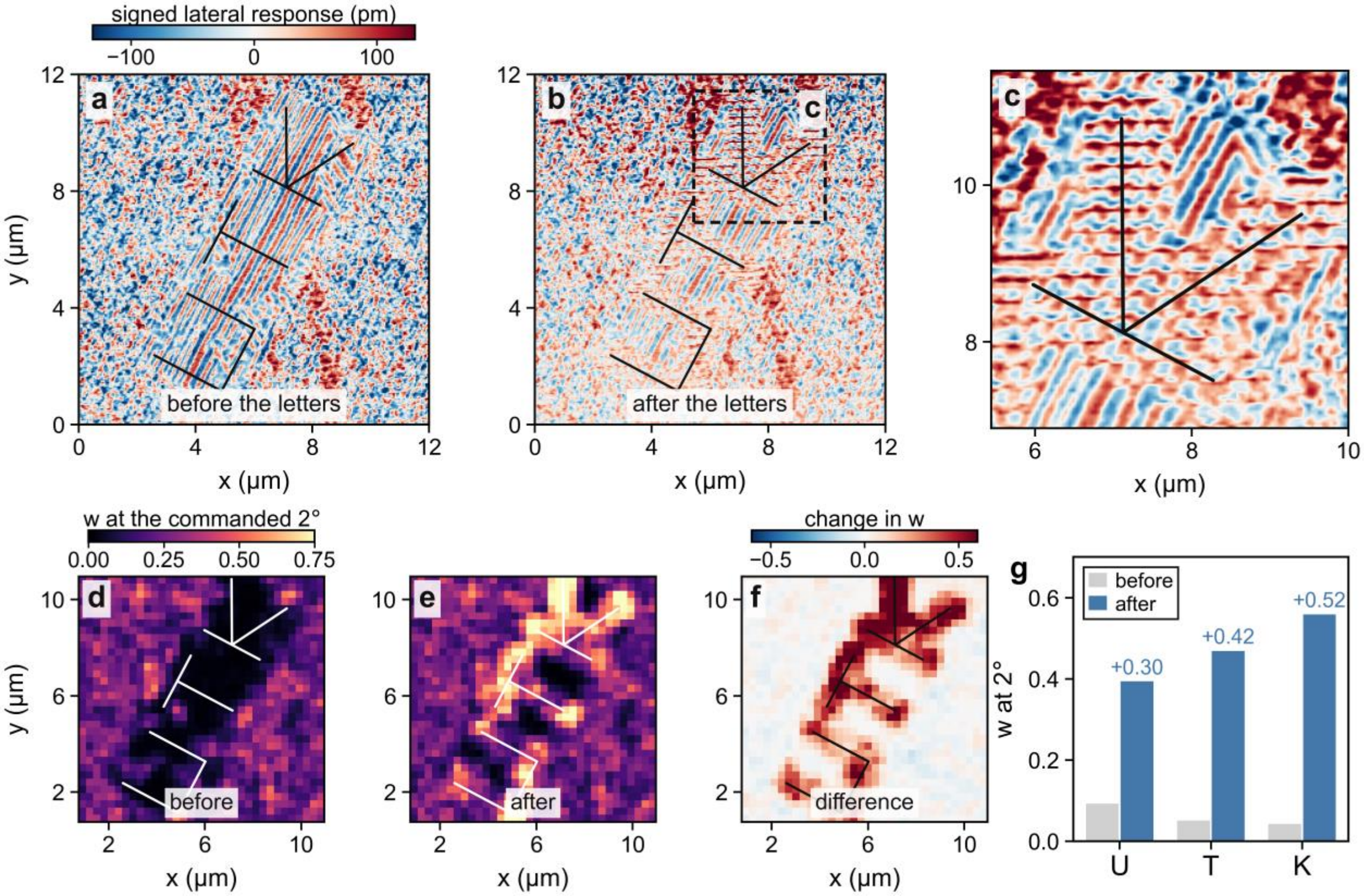

**Figure 5. Composing two rewriting rules to print UTK into the superdomain orientation (IT10b).** **a, b.** Signed lateral-PFM maps of the 12 $\mu m$ field before and after the letters, with the stroke centerlines overlaid. A raster along 62° had prepared the background, and an alternating-polarity lattice commanded to 2°, 60° away, was written only along the strokes with equal numbers of positive and negative sites. **c.** Detail of the K, in which the stripes on the strokes run along the command and the background stripes along the raster. **d–f.** Population of the commanded 2° family before and after the write, and its change, in which the mask is reproduced. **g.** Change on each letter relative to the spaces between the strokes. Each letter is about 3 $\mu m$ because a readout window must contain several lamellar periods. The out-of-plane response also changed during the raster, so the pattern is a measured in-plane response and not a complete polarization vector.

## 6. Discussion

### 6.1. Materials rules for in-plane superdomain control

The experiments establish a hierarchy among the tested control variables. The response remained confined to three crystallographic director families (Fig. S2a), and a population biased toward one family could be redirected toward another (Fig. 4 and Fig. S5e). For stationary pulse arrays, polarity alternation was necessary under the tested conditions (Fig. S5b), whereas exact matching between the array and lamellar periods was not (Fig. S5a). Directional selection also persisted below the exposure initially treated as critical (Fig. S5c). The spatial polarity pattern and initial superdomain state therefore had greater predictive value than simple commensurability or cumulative voltage–time exposure. The working protocol was a square lattice of 1 s, ±10 V pulses with rows spaced at half the local lamellar period, rotated to the target direction, and alternating in polarity between rows. It was effective on virgin film, on a region previously biased toward another direction, and on a raster-prepared background used for the UTK pattern in Fig. 5. These results do not identify the microscopic pathway. Mixed final states and the preparation-dependent raster response are compatible with contributions from nucleation, domain-wall motion, and elastic compatibility, which will require vector-resolved and time-resolved measurements to separate.

### 6.2. What the framework did

SPARC contributed to the campaign in three specific ways beyond command execution. First, operator feedback was converted into persistent experimental memory. Corrections concerning channel sign, estimator choice, same-location comparison, and positive controls were entered once in FINDINGS.md or PITFALLS.md and carried into later designs (Fig. S1). Second, the triage procedure in Fig. 2b separated instrument and analysis failures from changes in sample behavior: sideband inversion was recorded as an artifact in PITFALLS.md, whereas the preparation-dependent raster response superseded the earlier entry in FINDINGS.md and informed a later experiment. Third, selected pitfalls were converted into executable checks, allowing the agent to perform the two review steps in Campaign 2 (Fig. 6b). Across nine closed-loop cycles, the agent halted eight without operator intervention and rejected two provisional rules and revised a third based on new measurements before the final patterning experiment (Fig. S5f). The control software remained a weakness (Fig. 6d): six of eight launches failed because of faults that static checks could have detected, and the function intended to apply all 29 preflight checks was not called on the normal driver path. The observed autonomy therefore arose from the combination of persistent evidence, operator-defined constraints, and executable enforcement.

### 6.3. Defining and grounding the experiment

SPARC and Bayesian optimization address different stages of an experiment. Bayesian optimization is effective after the state, control variables, and objective have been defined. *Campaign 1* began before those choices were available: the state representation changed, the pulse-lattice operation emerged during the work, and the comparison used to evaluate an instrument action was revised. By the end of *Campaign 2*, the problem had been reduced to a small set of pulse-lattice parameters, a measured experimental floor, and a reproducible response range. Sequential optimization is appropriate at that stage.

This points to a limitation that we expect to be universal for agents working in open experimental environments. The critical bottleneck is the abstraction layer between LLM reasoning and the instrument. Agents that do mathematics work inside proof assistants in which every assumption and object is declared. Experimental agents need the equivalent for physical science: explicit definitions of the experimental state, instrument preconditions, permitted actions, provenance, controls, read-backs, uncertainty, and the range of conditions over which a conclusion is valid. A more mature experimental-agent architecture should make them part of the system from the beginning and make them visible to both the human operator and the agent.

The campaign also shows why physical grounding cannot stop at command generation. The system must verify that a command reached the instrument, produced the intended perturbation, yielded a valid measurement, and was interpreted with an observable tested on real instrument data. The key point is that these checks answer different questions and cannot substitute for one another. Knowing that a command is valid does not show that it was executed. Knowing that it was executed does not show that the intended physical change occurred. Observing a physical change does not guarantee that the measurement is reliable, and a reliable measurement does not automatically validate the quantity chosen to describe it.

### 6.4. Human-agent complementarity

The interaction record shows a complementary division of labor. The operator supplied instrument-specific knowledge, introduced several key physical insights, and challenged questionable interpretations. The agent rapidly implemented analysis routines, generated instrument-control programs and matched controls, maintained the evolving record, and later issued microscope commands directly. The language model itself did not change between campaigns. What changed was the shared experimental infrastructure: persistent memory, revised analysis code, explicit definitions of state and action, and controlled access to the microscope. Figure 6 summarizes this progression, and Table 2 lists the corresponding human and agent contributions.

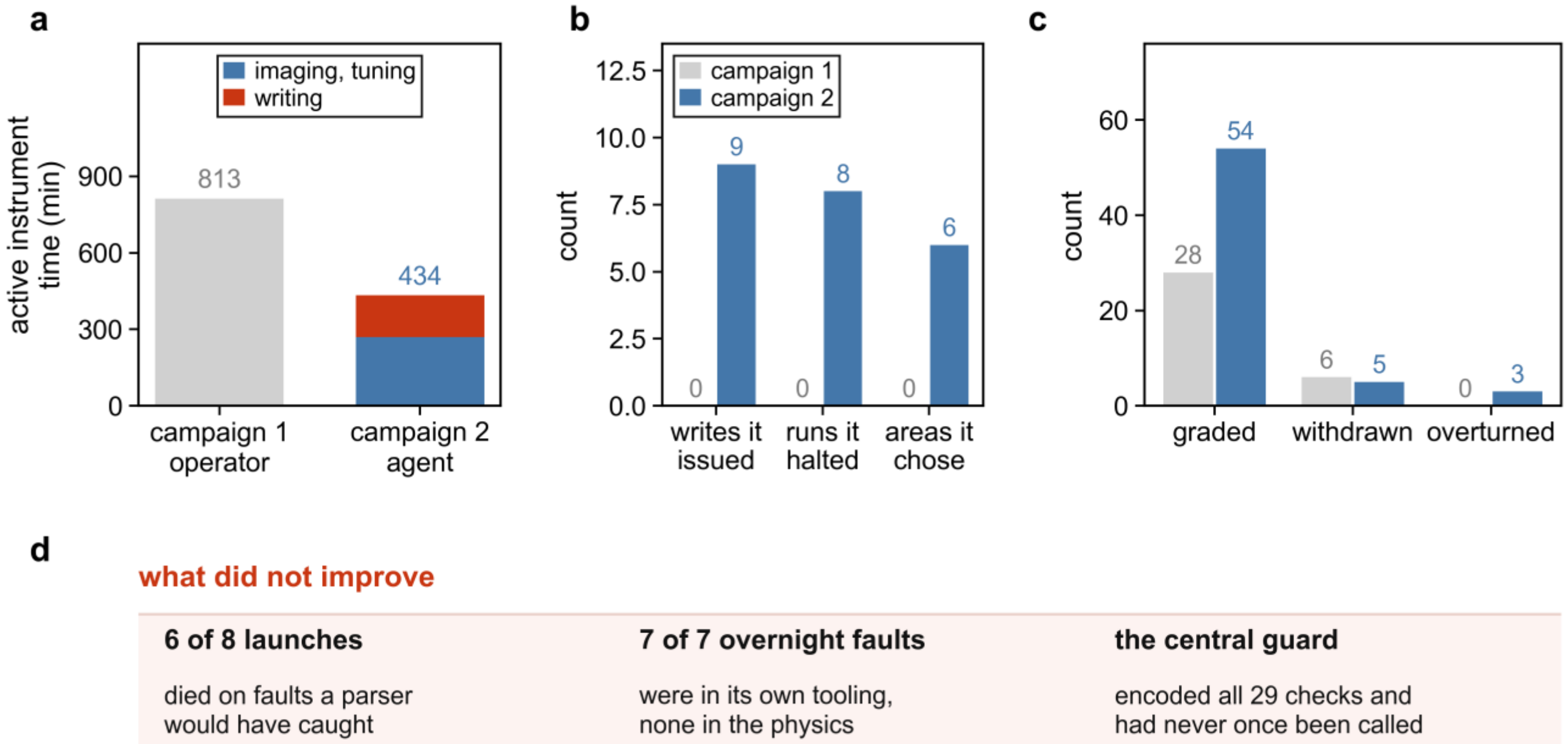

**Figure 6. Compact comparison of the two campaigns.** The operator-controlled campaign (*Campaign 1*) established the state representation, control primitive, and failure semantics. The agent-controlled campaign (*Campaign 2*) used those constructions to execute closed cycles, halt invalid runs, test its own rules, and expose failures in the agent-authored tooling. The axis-by-axis comparison and failure census are provided in Supplementary Section S12.

| Contribution | Agent | Operator |
|---|---|---|
| Theory and mechanism | Constructed the empirical state, symmetry arguments, operator classes, and quantitative control hypotheses; later tested and withdrew several of its own rules. | Reframed the scientific target, supplied apparatus physics, proposed point pulses, and challenged overconfident mechanism claims. |
| Experiment design | Generated parameterized geometries, controls, staged tests, abort criteria, and the autonomous drivers. | Selected or vetoed campaigns, introduced write families and geometry constraints, and set the physical objectives. |
| Measurement definition | Built the estimators, nulls, registration analysis, provenance checks, and synthetic validation. | Asked the questions that forced the principal revisions and performed post-campaign robustness checks. |
| Execution | No instrument commands in *Campaign 1*; nine closed cycles in *Campaign 2* under a programmatic wrapper. | Executed all *Campaign 1* writes and retained oversight and stop authority in *Campaign 2*. |

| Error detection and memory | Maintained graded findings and failure records, propagated fixes, and later self-halted invalid cycles. | Identified several literature, hardware, and experimental-logic errors and required graded persistent memory. |
|---|---|---|

**Table 2.** Complementary contributions of the agent and the operator. The full attribution audit, prompt coding, and limitations of retrospective attribution are provided in Supplementary Section S12.

### 6.5. Limitations and next experiments

The physical results carry the limitations of a first campaign. The main rewrite experiment in *Campaign 1* was demonstrated in a single region, and the autonomous set-reverse-hold experiments were performed on only a small number of independent areas. Retention was followed for no more than about an hour, and the written state was not always a single pure orientation. In the early pulse-lattice experiments, the command angle was also coupled to the number of pulse sites, and the lateral-PFM measurements were made with only one cantilever orientation. In addition, the mask-writing experiment produced a substantial change in the out-of-plane response. Resolving these limitations of the data will require replication in additional regions, matched write geometries, longer retention measurements, vector PFM or sample rotation, and measurements capable of distinguishing local switching from domain-wall motion and nucleation. A more complete experimental program is outlined (Supplementary Section S13).

The next stage should separate two questions that are related but scientifically distinct. The first is a control problem: once the geometry, exposure, and readout are sufficiently well defined, the pulse-lattice and raster operations can be treated as control primitives and optimized systematically. The second is a physics problem: what microscopic switching pathways produce the observed changes in domain orientation? Here the agent can be used differently, by maintaining several competing models of the switching process and selecting experiments that distinguish between them. Same-location vector-PFM measurements, interrupted writes, and direct tracking of domain-wall motion would help determine whether the observed response arises from local ferroelastic switching, nucleation and growth, wall motion, or combinations of these processes. Keeping the control and mechanism questions separate matters, because a successful writing strategy does not by itself establish the microscopic physics that makes it work.

## 7. Summary

Many microscopy studies begin before the experimental problem can be expressed in terms of fixed variables, available operations, and a quantitative objective. In this setting, Bayesian optimization and model-based planning are premature because the relevant observable and effective control protocols are still being established. The number of tests may also be constrained by limited sample area and microscope time. We developed SPARC for this early stage of experimentation by coupling human operators with a coding agent. The agent analyzed PFM data, proposed experiments, and later controlled the microscope after appropriate safeguards had been implemented. Experimental conclusions, together with their confidence and applicable conditions, were recorded in FINDINGS.md. Known failures and the checks were recorded in PITFALLS.md. The agent read both records when planning each subsequent experiment.

Applied to (111)-oriented $PbZr_{0.2}Ti_{0.8}O_3$ film, this workflow identified practical approaches for controlling the in-plane superdomain population. During *Campaign 1*, the agent analyzed earlier PFM measurements, developed a quantitative descriptor of the superdomain configuration, and designed the alternating-polarity pulse-lattice protocol. The operator retained physical control of the microscope during this stage. During *Campaign 2*, the agent operated the instrument directly, tested the assumptions carried forward from the first campaign, and revised them based on new measurements. The pulse lattice preferentially increased the population of a targeted crystallographic orientation and redirected previously written regions toward another allowed orientation. The reconfigured orientation persisted during repeated PFM imaging over tens of minutes, although longer-term stability was not tested. Within the investigated range, polarity alternation was important for directional selection, whereas exact period matching and the exposure threshold were not required. A bipolar raster also aligned previously unwritten regions along the scan direction. Combining raster preparation with masked pulse-lattice writing produced a UTK-shaped pattern in the measured in-plane response.

The transition to direct instrument control also revealed important practical limitations. A software command could be issued correctly without reaching the microscope or producing the intended physical actions. Therefore, reliable operation required separate checks for command execution, sample response, control measurements, data provenance, and the validity of image-derived descriptors. Although several safeguards presented in this work remained specific to the microscope and measurement protocol, SPARC provided a framework to implement safeguards on other instruments with prior knowledge and human supervision.

Automated instruments can acquire data and execute operations faster than a researcher can interpret each result and design the next measurement.[42] SPARC can shorten this experimental loop by supporting data analysis, record-keeping, and selection of subsequent measurements. The researcher nevertheless remains responsible for defining the scientific objective, evaluating physical interpretations, and reformulating the problem when new observations challenge previous assumptions. SPARC therefore complements conventional automation and sequential optimization by helping convert an open-ended investigation into a well-defined experimental problem. Once the observable, control space, and objective have stabilized, the resulting problem can be transferred to conventional sequential optimization.

## 8. Materials and methods

### 8.1. Sample, imaging, and state representation

As a model system, we used a (111)-oriented $PbZr_{0.2}Ti_{0.8}O_3$ film and a metal-coated probe. Heterostructures consisting of 150 nm $PbZr_{0.2}Ti_{0.8}O_3$ (PZTO) and 30 nm $La_{0.67}Sr_{0.33}MnO_3$ (LSMO) were grown on $SrTiO_3$ (STO) substrates (111) orientation (MTI Corp.) using pulsed-laser deposition (PLD). A KrF excimer laser (λ = 248 nm, LPX 300, Coherent) was employed to ablate ceramic targets with nominal compositions of $Pb_{1.2}Zr_{0.2}Ti_{0.8}O_3$ and $La_{0.67}Sr_{0.33}MnO_3$ (Praxair Inc.). Both the PZT and LSMO layers were deposited under identical conditions: a target-to-substrate distance of 60 mm, a substrate heater temperature of 635 °C, an oxygen partial pressure of 200 mTorr, a laser fluence of 1.5 J $cm^{-2}$, and a laser repetition rate of 4 Hz. After deposition, the samples were cooled to room temperature at a rate of 10 °C $min^{-1}$ in flowing oxygen at a pressure of ~700 Torr.

Principal images were 8 $\mu m$ frames sampled at 256 pixels and acquired in lateral dual-frequency resonance tracking (DART), with the cantilever long axis along the image x direction, so that a positive signed lateral response corresponds to torsion in the +x sense. Writes used a metal-coated probe at a tip bias of 10 V, a scan angle set by the commanded director, a line pitch of one half of the locally measured lamellar period, and a tip velocity of 0.5 $\mu m$ per second unless stated otherwise. The contact resonances, the frequencies at which the tip-sample contact vibrates most strongly and which DART tracks so that the signal is not lost when the tip changes, were 650 kHz lateral and 350 kHz vertical, retuned before every baseline frame. The local stripe-director triad was fitted as a rigid three-member set separated by 60°. Signed lateral response maps were constructed from amplitude and phase channels, with a sideband-consistency guard added after anti-correlated channels were identified. Angular power was calculated from the two-dimensional Fourier spectrum over the analyzed panel cores and integrated within +/-15° of target directions. No-treatment image pairs provided a run-specific temporal floor. Full channel provenance, estimator definitions, and the synthetic-to-real transfer test are given in Supplementary Sections S4.2, S7, S8, and S14. Vector-PFM interpretation, dual-frequency resonance tracking, band-excitation acquisition, and image registration follow established methods.[43-46] The physical priors for ferroelastic switching and (111)-oriented superdomains were drawn from earlier experimental and theoretical studies.[41, 47-49]

### 8.2. Pulse-lattice and raster writing

The pulse lattice consisted of stationary biased sites on a square grid rotated to the commanded director. In the principal region, a representative 280 nm lamellar scale gave a nominal 140 nm row spacing. *Campaign 1* write A used 14 x 14 sites per addressed 2 $\mu m$ panel and write B used 12 x 12 because the rotated bounding box reduced the number of sites that fit. Each site received a nominal 10 V for 1 s with alternating polarity between rows and equal counts of positive and negative voltage-time exposure within a panel. Current was not measured, so all dose quantities are voltage-time exposures rather than electrical charge. *Campaign 2* used the same family of pulse-lattice operations together with large-area raster preparation and mask-defined lattices. Detailed geometries and write programs are in Supplementary Sections S7, S11, and S14.2.

### 8.3. Agent system and interaction record

The agent was Claude Opus 4.1 running inside Claude Code on the instrument-control computer, with shell and file tools and no direct network access to the vendor software. The same model and tool set were used in both campaigns. Persistent state was kept in the shared notebook, in FINDINGS.md and PITFALLS.md (Supplementary Section S6), and, in *Campaign 2*, in a JSON campaign state file. *Campaign 1* was human mediated: the agent generated analyses, experiment cells, and write programs, and the operator executed the instrument actions. In *Campaign 2* the agent issued commands through a Python wrapper while the operator remained present. The complete transcript, prompt coding, tool-call counts, and chronology are described in Supplementary Sections S5 and S14.4 and will be released with the data. The system design also drew on prior work on automated microscopy, reward-based image analysis, and coding agents for scientific discovery.[50-52]

### 8.4. Autonomous loop, provenance, and statistical interpretation

Each direct-actuation iteration followed propose, preflight, place, write, read, analyze, log, and decide steps. Instrument automation was realized through the Automated Experiment in Scanning Probe Microscopy (AESPM) control interface.[32] An exclusive lock prevented two processes from addressing the instrument simultaneously, and a file-level stop command could halt the loop before a new instrument action. Candidate writes were constrained by scanner range, bias and exposure limits, declared footprints, and campaign budgets. The detailed distinction between checks that executed in the driver and checks that existed only in auxiliary code is given in Supplementary Section S9. Experimental provenance was reconstructed from file headers, trajectory files, timestamps, resonance channels, and frame order. The manuscript reports deterministic contrasts against measured no-treatment floors rather than population-level inference across many independent sample areas. Spatially resolved pattern statistics are descriptive. Publication-quality significance for spatial windows should use block resampling to account for correlation.

## Data availability

The conversational transcript, raw piezoresponse frames, trajectory files, agent-maintained findings and failure records, tagged notebook, interaction-coding scheme, and analysis scripts required to reproduce the reported quantities are deposited at
https://github.com/RichardLiuCoding/Publications/tree/main/SPARC-PZTO111

## Acknowledgements

This work was supported by the National Science Foundation through the Ceramics program NSF 2523284. C.-C.L. acknowledges that this research was sponsored by the Army Research Laboratory and was accomplished under Cooperative Agreement Number W911NF-24-2-0100. The views and conclusions contained in this document are those of the authors and should not be interpreted as representing the official policies, either expressed or implied, of the Army Research Laboratory or the U.S. Government. The U.S. Government is authorized to reproduce and distribute reprints for Government purposes notwithstanding any copyright notation herein. J.K. acknowledges that this material is based upon work supported by the Air Force Office of Scientific Research under award number FA9550-24-1-0266. Any opinions, findings, and conclusions or recommendations expressed in this material are those of the author(s) and do not necessarily reflect the views of the United States Air Force. L.W.M. acknowledges the support of the NSF PCL-Test Bed READINESS program under grant no. 2607553.